\documentclass[titlepage,12pt]{article}
\usepackage{amsmath}
\usepackage[dvips]{graphicx}
\usepackage{subfigure}

\begin{document}

\title{Bound states of the Duffin-Kemmer-Petiau equation with a mixed
minimal-nonminimal vector cusp potential}
\date{}
\author{A.S. de Castro\thanks{%
castro@pq.cnpq.br.} \\
%EndAName
\\
UNESP - Campus de Guaratinguet\'{a}\\
Departamento de F\'{\i}sica e Qu\'{\i}mica\\
12516-410 Guaratinguet\'{a} SP - Brazil}
\date{ }
\maketitle

\begin{abstract}
The problem of spin-0 and spin-1 bosons subject to a general mixing of
minimal and nonminimal vector cusp potentials is explored in a unified way
in the context of the Duffin-Kemmer-Petiau theory. Effects on the
bound-state solutions due to a short-range interaction are discussed in some
detail. \newline
\newline
\newline
\newline
Keywords: DKP equation, Klein's paradox, pair production, nonminimal
coupling \newline
\newline
PACS Numbers: 03.65.Ge, 03.65.Pm
\end{abstract}

\section{Introduction}

The first-order Duffin-Kemmer-Petiau (DKP) formalism \cite{pet}-\cite{kem}
describes spin-0 and spin-1 particles and has been used to analyze
relativistic interactions of spin-0 and spin-1 hadrons with nuclei as an
alternative to their conventional second-order Klein-Gordon and Proca
counterparts. The onus of equivalence between the formalisms represented an
objection to the DKP theory for a long time and only recently it was shown
that they yield the same results in the case of minimally coupled vector
interactions, on the condition that one correctly interprets the components
of the DKP spinor \cite{now}-\cite{lun}. However, the equivalence between
the DKP and the Proca formalisms has already a precedent \cite{mr}. The
equivalence does not maintain if one considers partially conserved currents
\cite{FIS}. Furthermore, the DKP formalism enjoys a richness of couplings
not capable of being expressed in the Klein-Gordon and Proca theories.

Nonminimal vector potentials, added by other kinds of Lorentz structures,
have already been used successfully in a phenomenological context for
describing the scattering of mesons by nuclei \cite{cla1}-\cite{cla2}.
Nonminimal vector coupling with a quadratic potential \cite{Ait}, with a
linear potential \cite{kuli}, and mixed space and time components with a
step potential \cite{ccc3}-\cite{ccc2} and a linear potential \cite{jpa}
have been explored in the literature. See also Ref. \cite{jpa} for a
comprehensive list of references on other sorts of couplings and functional
forms for the potential functions. In Ref. \cite{jpa} it was shown that when
the space component of the coupling is stronger than its time component the
linear potential, a sort of vector DKP oscillator, can be used as a model
for confining bosons.

The cusp potential in the form $e^{-|x|/\lambda }$ (screened Coulomb
potential in a two-dimensional space-time world) \thinspace has been
analyzed and its analytical solutions have been found for the Dirac equation
with vector \cite{ada}-\cite{vil1}, scalar \cite{kag3} and pseudoscalar \cite%
{asc} couplings, and for the Klein-Gordon equation with vector \cite{kag1}-%
\cite{vil3} and scalar \cite{kag2} couplings, and a mixed scalar-vector
coupling \cite{ijmpa}. As has been emphasized in Refs. \cite{kag1} and \cite%
{kag2}, the solutions of relativistic equations with this sort of potential
may find applications in the study of pionic atoms, doped Mott insulators,
doped semiconductors, interaction between ions, quantum dots surrounded by a
dielectric or a conducting medium, protein structures, etc. The bound states
for the Klein-Gordon equation with a minimal vector coupling differ
radically from those ones for the Dirac equation. For an enough deep and
narrow potential the Klein-Gordon equation exhibits the phenomenon called
Schiff-Snyder-Weinberg effect (SSWE) \cite{ssw}. Such an effect manifests by
additional antiparticle bound states in a potential attractive only for
particles. For critical depths, the particle and antiparticle energy levels
coalesce and there unveils a new channel for the pair production. Popov \cite%
{pop} argued that the SSWE is inherent to short-range interactions and
should not be expected for large-range potentials. Nevertheless, Klein and
Rafelski \cite{kr} used a purported SSWE in a Coulomb potential for
speculating about the Bose condensation and the stability of extremely high
atomic number nuclei and, right away, were severely criticized \cite{baw}.
As a matter of fact, the investigation of the bound-state solutions of the
Klein-Gordon equation with different functional forms for the potential
validates Popov's conjecture \cite{vil2}-\cite{vil3},\cite{flei}-\cite{roj}.

The main purpose of the present article is to report on the properties of
the DKP theory with time components of minimal and nonminimal vector cusp
potentials for spin-0 and spin-1 bosons in a unified way. This sort of
mixing, beyond its potential physical applications, shows to be a powerful
tool to obtain a deeper insight about the nature of the DKP equation and its
solutions as far as it explores the differences between minimal and
nonminimal couplings. The problem is mapped into an exactly solvable
Sturm-Liouville problem of a Schr\"{o}dinger-like equation with an effective
symmetric Morse-like potential, or an effective cusp potential in the
particular circumstance of a pure minimal coupling.

\section{Vector couplings in the DKP equation}

The DKP equation for a free boson is given by \cite{kem} (with units in
which $\hbar =c=1$)%
\begin{equation}
\left( i\beta ^{\mu }\partial _{\mu }-m\right) \psi =0  \label{dkp}
\end{equation}%
\noindent where the matrices $\beta ^{\mu }$\ satisfy the algebra $\beta
^{\mu }\beta ^{\nu }\beta ^{\lambda }+\beta ^{\lambda }\beta ^{\nu }\beta
^{\mu }=g^{\mu \nu }\beta ^{\lambda }+g^{\lambda \nu }\beta ^{\mu }$
\noindent and the metric tensor is $g^{\mu \nu }=\,$diag$\,(1,-1,-1,-1)$.
That algebra generates a set of 126 independent matrices whose irreducible
representations are a trivial representation, a five-dimensional
representation describing the spin-0 particles and a ten-dimensional
representation associated to spin-1 particles. The second-order Klein-Gordon
and Proca equations are obtained when one selects the spin-0 and spin-1
sectors of the DKP theory. A well-known conserved four-current is given by $%
J^{\mu }=\bar{\psi}\beta ^{\mu }\psi $\noindent $/2$ where the adjoint
spinor $\bar{\psi}$ is given by $\bar{\psi}=\psi ^{\dagger }\eta ^{0}$ with $%
\eta ^{0}=2\beta ^{0}\beta ^{0}-1$, provided $\beta ^{\mu }=\eta ^{0}\left(
\beta ^{\mu }\right) ^{\dagger }\eta ^{0}$. The time component of this
current is not positive definite but it may be interpreted as a charge
density. Then the normalization condition $\int d\tau \,J^{0}=\pm 1$ can be
expressed as%
\begin{equation}
\int d\tau \,\bar{\psi}\beta ^{0}\psi =\pm 2  \label{norm}
\end{equation}%
where the plus (minus) sign must be used for a positive (negative) charge.

With the introduction of interactions, the DKP equation can be written as%
\begin{equation}
\left( \beta ^{\mu }p_{\mu }-m-V\right) \psi =0
\end{equation}%
and $J^{\mu }$ satisfies the equation \cite{jpa}%
\begin{equation}
\partial _{\mu }J^{\mu }+\frac{i}{2}\bar{\psi}\left( V-\eta ^{0}V^{\dagger
}\eta ^{0}\right) \psi =0  \label{corrent2}
\end{equation}%
Thus, if $V$ is Hermitian with respect to $\eta ^{0}$ then the four-current
will be conserved. The more general potential matrix $V$ is written in terms
of 25 (100) linearly independent matrices pertinent to the
five(ten)-dimensional irreducible representation associated to the scalar
(vector) sector and can be written in terms of well-defined Lorentz
structures. For the spin-0 sector there are two scalar, two vector and two
tensor terms \cite{gue}, whereas for the spin-1 sector there are two scalar,
two vector, a pseudoscalar, two pseudovector and eight tensor terms \cite%
{vij}. Considering only the vector terms, $V$ is in the form \cite{gue},
\cite{ume}%
\begin{equation}
\left( i\beta ^{\mu }\partial _{\mu }-m-\beta ^{\mu }A_{\mu }^{\left(
1\right) }-i[P,\beta ^{\mu }]A_{\mu }^{\left( 2\right) }\right) \psi =0
\label{dkp2}
\end{equation}%
where $P$ and $[P,\beta ^{\mu }]$ are independent elements of the DKP subalgebra
obtained by simple products of the $\beta ^{\mu }$ matrices plus the unit
matrix in such a way that $\bar{\psi}[P,\beta ^{\mu }]\psi $ behaves like a
vector under a Lorentz transformation as does $\bar{\psi}\beta ^{\mu }\psi $%
. Once again $\partial _{\mu }J^{\mu }=0$, provided $P=\eta ^{0}\left(
P\right) ^{\dagger }\eta ^{0}$ and $i[P,\beta ^{\mu }]=\eta ^{0}\left(
i[P,\beta ^{\mu }]\right) ^{\dagger }\eta ^{0}$. Notice that the vector
potential $A_{\mu }^{\left( 1\right) }$ is minimally coupled but not $A_{\mu
}^{\left( 2\right) }$. If the terms in the potentials $A_{\mu }^{\left(
1\right) }$ and $A_{\mu }^{\left( 2\right) }$ are time-independent one can
write $\psi (\vec{r},t)=\phi (\vec{r})\exp (-iEt)$, where $E$ is the energy
of the boson, in such a way that the time-independent DKP equation becomes%
\begin{equation}
\left[ \beta ^{0}\left( E-A_{0}^{\left( 1\right) }\right) +i\beta ^{i}\left(
\partial _{i}+iA_{i}^{\left( 1\right) }\right) -\left( m+i[P,\beta ^{\mu
}]A_{\mu }^{\left( 2\right) }\right) \right] \phi =0  \label{DKP10}
\end{equation}%
In this case \ $J^{\mu }=\bar{\phi}\beta ^{\mu }\phi /2$ does not depend on
time, so that the spinor $\phi $ describes a stationary state. Note that the
time-independent DKP equation is invariant under a simultaneous shift of $E$
and $A_{0}^{\left( 1\right) }$, such as in the Schr\"{o}dinger equation, but
the invariance does not maintain regarding $E$ and $A_{0}^{\left( 2\right) }$%
. It can be shown (see Ref. \cite{jpa}) that any two stationary states with
distinct energies and subject to the boundary conditions%
\begin{equation}
\int d\tau \,\partial _{i}\left( \bar{\phi}_{k}\beta ^{i}\phi _{k^{\prime
}}\right) =0
\end{equation}%
are orthogonal in the sense that $\int d\tau \,\bar{\phi}_{k}\beta ^{0}\phi
_{k^{\prime }}=0$, for $E_{k}\neq E_{k^{\prime }}$. In addition, in view of (%
\ref{norm}) the spinors $\phi _{k}$ and $\phi _{k^{\prime }}$ are said to be
orthonormal if%
\begin{equation}
\int d\tau \,\bar{\phi}_{k}\beta ^{0}\phi _{k^{\prime }}=\pm 2\delta
_{E_{k}E_{k^{\prime }}}  \label{orto8}
\end{equation}

The charge-conjugation operation changes the sign of the minimal interaction
potential,\ i.e.\textit{\ }changes the sign of \ $A_{\mu }^{\left( 1\right)
} $. This can be accomplished by the transformation $\psi \rightarrow \psi
_{c}=\mathcal{C}\psi =CK\psi $, where $K$ denotes the complex conjugation
and $C$ is a unitary matrix such that $C\beta ^{\mu }=-\beta ^{\mu }C$. The
matrix that satisfies this relation is $C=\exp \left( i\delta _{C}\right)
\eta ^{0}\eta ^{1}$. The phase factor $\exp \left( i\delta _{C}\right) $ is
equal to $\pm 1$, thus $E\rightarrow -E$. Note also that $J^{\mu
}\rightarrow -J^{\mu }$, as should be expected for a charge current.
Meanwhile $C$ anticommutes with $[P,\beta ^{\mu }]$ and the
charge-conjugation operation entails no change on $A_{\mu }^{\left( 2\right)
}$. The invariance of the nonminimal vector potential under charge
conjugation means that it does not couple to the charge of the boson. In
other words, $A_{\mu }^{\left( 2\right) }$ does not distinguish particles
from antiparticles. Hence, whether one considers spin-0 or spin-1 bosons,
this sort of interaction can not exhibit Klein's paradox.

For the case of spin 0, we use the representation for the $\beta ^{\mu }$\
matrices given by \cite{ned1}%
\begin{equation}
\beta ^{0}=%
\begin{pmatrix}
\theta & \overline{0} \\
\overline{0}^{T} & \mathbf{0}%
\end{pmatrix}%
,\quad \beta ^{i}=%
\begin{pmatrix}
\widetilde{0} & \rho _{i} \\
-\rho _{i}^{T} & \mathbf{0}%
\end{pmatrix}%
,\quad i=1,2,3  \label{rep}
\end{equation}%
\noindent where%
\begin{eqnarray}
\ \theta &=&%
\begin{pmatrix}
0 & 1 \\
1 & 0%
\end{pmatrix}%
,\quad \rho _{1}=%
\begin{pmatrix}
-1 & 0 & 0 \\
0 & 0 & 0%
\end{pmatrix}
\notag \\
&&  \label{rep2} \\
\rho _{2} &=&%
\begin{pmatrix}
0 & -1 & 0 \\
0 & 0 & 0%
\end{pmatrix}%
,\quad \rho _{3}=%
\begin{pmatrix}
0 & 0 & -1 \\
0 & 0 & 0%
\end{pmatrix}
\notag
\end{eqnarray}%
\noindent $\overline{0}$, $\widetilde{0}$ and $\mathbf{0}$ are 2$\times $3, 2%
$\times $2 and 3$\times $3 zero matrices, respectively, while the
superscript T designates matrix transposition. Here the matrix $P$ appearing
in (\ref{dkp2}) can be written as \cite{gue} $P=\left( \beta ^{\mu }\beta
_{\mu }-1\right) /3=\mathrm{diag}\,(1,0,0,0,0)$. In this case $P$ picks out
the first component of the DKP spinor. The five-component spinor can be
written as $\psi ^{T}=\left( \psi _{1},...,\psi _{5}\right) $ in such a way
that the time-independent DKP equation for a boson constrained to move along
the $X$-axis, restricting ourselves to time-like components of
four-dimensional vector potentials ($\vec{A}^{\left( 1\right) }=\vec{A}%
^{\left( 2\right) }=\vec{0}$) depending only on $x$, decomposes into%
\begin{equation*}
\left( \frac{d^{2}}{dx^{2}}+k^{2}\right) \phi _{1}=0
\end{equation*}%
\begin{equation}
\phi _{2}=\frac{1}{m}\left( E-A_{0}^{\left( 1\right) }+iA_{0}^{\left(
2\right) }\right) \,\phi _{1}  \label{dkp4}
\end{equation}%
\begin{equation*}
\phi _{3}=\frac{i}{m}\frac{d}{dx}\phi _{1},\quad \phi _{4}=\phi _{5}=0
\end{equation*}%
where%
\begin{equation}
k^{2}=\left( E-A_{0}^{\left( 1\right) }\right) ^{2}-m^{2}+\left(
A_{0}^{\left( 2\right) }\right) ^{2}  \label{k}
\end{equation}%
Meanwhile,
\begin{equation}
J^{0}=\frac{E-A_{0}^{\left( 1\right) }}{m}\,|\phi _{1}|^{2},\quad J^{1}=%
\frac{1}{m}\text{Im}\left( \phi _{1}^{\ast }\,\frac{d\phi _{1}}{dx}\right)
\label{corrente4}
\end{equation}%
It is worthwhile to note that $J^{0}$ becomes negative in regions of space
where $E<A_{0}^{\left( 1\right) }$ (a circumstance associated to Klein's
paradox) and that $A_{\mu }^{\left( 2\right) }$ does not intervene
explicitly in the current. The orthonormalization formula (\ref{orto8})
becomes%
\begin{equation}
\int\limits_{-\infty }^{+\infty }dx\,\,\frac{\frac{E_{k}+E_{k^{\prime }}}{2}%
-A_{0}^{\left( 1\right) }}{m}\,\phi _{1k}^{\ast }\phi _{1k^{\prime }}=\pm
\delta _{E_{k}E_{k^{\prime }}}  \label{ORTO1}
\end{equation}%
regardless $A_{\mu }^{\left( 2\right) }$. Eq. (\ref{ORTO1}) is in agreement
with the orthonormalization formula for the Klein-Gordon theory in the
presence of a minimally coupled potential \cite{puk}. This is not
surprising, because, after all, both DKP equation and Klein-Gordon equation
are equivalent under minimal coupling.

For the case of spin 1, the $\beta ^{\mu }$\ matrices are \cite{ned2}%
\begin{equation*}
\beta ^{0}=%
\begin{pmatrix}
0 & \overline{0} & \overline{0} & \overline{0} \\
\overline{0}^{T} & \mathbf{0} & \mathbf{I} & \mathbf{0} \\
\overline{0}^{T} & \mathbf{I} & \mathbf{0} & \mathbf{0} \\
\overline{0}^{T} & \mathbf{0} & \mathbf{0} & \mathbf{0}%
\end{pmatrix}%
\end{equation*}%
\begin{equation}
\beta ^{i}=%
\begin{pmatrix}
0 & \overline{0} & e_{i} & \overline{0} \\
\overline{0}^{T} & \mathbf{0} & \mathbf{0} & -is_{i} \\
-e_{i}^{T} & \mathbf{0} & \mathbf{0} & \mathbf{0} \\
\overline{0}^{T} & -is_{i} & \mathbf{0} & \mathbf{0}%
\end{pmatrix}
\label{betaspin1}
\end{equation}%
\noindent where $s_{i}$ are the 3$\times $3 spin-1 matrices $\left(
s_{i}\right) _{jk}=-i\varepsilon _{ijk}$, $e_{i}$ are the 1$\times $3
matrices $\left( e_{i}\right) _{1j}=\delta _{ij}$ and $\overline{0}=%
\begin{pmatrix}
0 & 0 & 0%
\end{pmatrix}%
$, while\textbf{\ }$\mathbf{I}$ and $\mathbf{0}$\textbf{\ }designate the 3$%
\times $3 unit and zero matrices, respectively. In this representation $%
P=\,\beta ^{\mu }\beta _{\mu }-2=\mathrm{diag}\,(1,1,1,1,0,0,0,0,0,0)$, i.e.
$P$ projects out the four upper components of the DKP spinor. \noindent With
the spinor written as $\psi ^{T}=\left( \psi _{1},...,\psi _{10}\right) $,
and partitioned as%
\begin{equation*}
\psi _{I}^{\left( +\right) }=\left(
\begin{array}{c}
\psi _{3} \\
\psi _{4}%
\end{array}%
\right) ,\quad \psi _{I}^{\left( -\right) }=\psi _{5}
\end{equation*}%
\begin{equation}
\psi _{II}^{\left( +\right) }=\left(
\begin{array}{c}
\psi _{6} \\
\psi _{7}%
\end{array}%
\right) ,\quad \psi _{II}^{\left( -\right) }=\psi _{2}  \label{part}
\end{equation}%
\begin{equation*}
\psi _{III}^{\left( +\right) }=\left(
\begin{array}{c}
\psi _{10} \\
-\psi _{9}%
\end{array}%
\right) ,\quad \psi _{III}^{\left( -\right) }=\psi _{1}
\end{equation*}%
the one-dimensional time-independent DKP equation with time-like components
of vector potentials can be expressed as
\begin{equation*}
\left( \frac{d^{2}}{dx^{2}}+k^{2}\right) \phi _{I}^{\left( \pm \right) }=0
\end{equation*}%
\begin{equation}
\phi _{II}^{\left( \pm \right) }=\frac{1}{m}\left( E-A_{0}^{\left( 1\right)
}\pm iA_{0}^{\left( 2\right) }\right) \,\phi _{I}^{\left( \pm \right) }
\label{spin1-ti}
\end{equation}%
\begin{equation*}
\phi _{III}^{\left( \pm \right) }=\frac{i}{m}\frac{d}{dx}\phi _{I}^{\left(
\pm \right) },\quad \phi _{8}=0
\end{equation*}%
where $k$ is again given by (\ref{k}). Now the components of the
four-current are%
\begin{equation*}
J^{0}=\frac{E-A_{0}^{\left( 1\right) }}{m}\left( |\phi _{I}^{\left( +\right)
}|^{2}+|\phi _{I}^{\left( -\right) }|^{2}\right)
\end{equation*}%
\begin{equation}
J^{1}=\frac{1}{m}\text{Im}\left( \phi _{I}^{\left( +\right) \dagger }\,\frac{%
d\phi _{I}^{\left( +\right) }}{dx}+\phi _{I}^{\left( -\right) \dagger }\,%
\frac{d\phi _{I}^{\left( -\right) }}{dx}\right)  \label{CUR2}
\end{equation}%
and the orthonormalization expression (\ref{orto8}) takes the form%
\begin{equation}
\int\limits_{-\infty }^{+\infty }dx\,\,\frac{\frac{E_{k}+E_{k^{\prime }}}{2}%
-A_{0}^{\left( 1\right) }}{m}\,\left( \phi _{Ik}^{\left( +\right) \dagger
}\phi _{Ik^{\prime }}^{\left( +\right) }+\phi _{Ik}^{\left( -\right) \dagger
}\phi _{Ik^{\prime }}^{\left( -\right) }\right) =\pm \delta
_{E_{k}E_{k^{\prime }}}
\end{equation}%
Just as for scalar bosons, $J^{0}<0$ for $E<A_{0}^{\left( 1\right) }$ and $%
A_{\mu }^{\left( 2\right) }$ does not appear in the current. Similarly, $%
A_{\mu }^{\left( 2\right) }$ do not manifest explicitly in the
orthonormalization formula.

Comparison between the two sets of formulas for the spin-0 and spin-1
sectors of the DKP theory evidences that vector bosons and scalar bosons
behave in a similar way.

\section{The cusp potential}

Now we are in a position to use the DKP equation with specific forms for
vector interactions. Let us focus our attention on time components of
minimal and nonminimal vector potentials in the form of a cusp potential
\begin{equation}
A=-V_{0}\exp \left( -\frac{|x|}{\lambda }\right) ,\quad V_{0}>0  \label{sc1}
\end{equation}%
\noindent with
\begin{equation}
A_{0}^{\left( 1\right) }=g_{1}A,\quad A_{0}^{\left( 2\right) }=g_{2}A
\label{sc1a}
\end{equation}%
where the coupling constants, $g_{1}$ and $g_{2}$, are dimensionless real
parameters and $\lambda $, related to the range of the interaction, is a
positive parameter. In this case the first equations of (\ref{dkp4}) and (%
\ref{spin1-ti}) transmutes into%
\begin{equation}
-\frac{1}{2m}\frac{d^{2}\Phi }{dx^{2}}+V_{\mathtt{eff}}\,\Phi =E_{\mathtt{eff%
}}\,\Phi  \label{sc1b}
\end{equation}%
where $\Phi $ is equal to $\phi _{1}$ for the scalar sector, and to $\phi
_{I}^{\left( \pm \right) }$ for the vector sector, with%
\begin{equation*}
V_{\mathtt{eff}}=V_{1}\exp \left( -\frac{|x|}{\lambda }\right) +V_{2}\exp
\left( -2\frac{|x|}{\lambda }\right)
\end{equation*}
\begin{equation}
E_{\mathtt{eff}}=\frac{E^{2}-m^{2}}{2m}  \label{sc2}
\end{equation}%
and
\begin{equation}
V_{1}=-\frac{Eg_{1}V_{0}}{m},\quad V_{2}=-\frac{g_{1}^{2}+g_{2}^{2}}{2m}%
\,V_{0}^{2}  \label{sc3}
\end{equation}

\noindent Therefore, one has to search for bounded solutions in an effective
symmetric Morse-like potential for $g_{1}\neq 0$, or in a cusp potential for
the case of a pure nonminimal vector potential ($g_{1}=0$). Note carefully
that the DKP equation can furnish a discrete spectrum only if%
\begin{equation}
V_{1}<|V_{2}|  \label{sc4}
\end{equation}%
This is so because the effective potential approaches zero as $%
|x|\rightarrow \infty $ and has a minimum value equal to $V_{1}-|V_{2}|$.
Only in this circumstance the effective potential presents a potential-well
structure permitting bounded solutions in the range $|E|<m$ ($E_{\mathtt{eff}%
}<0$). The energies in the range $|E|>m$ correspond to the continuum. Also
note that the spectrum changes sign under the transformation $%
g_{1}\rightarrow -g_{1}$, but it remains invariant under $g_{2}\rightarrow
-g_{2}$. The DKP energies are obtained by inserting the effective
eigenvalues into (\ref{sc2}). When $g_{1}\neq 0$ the effective potential
depends on the energy and the single-well potential is deeper and larger for
one sign of energy than for the other one. When $g_{1}>0$, for instance, the
single-well potential is deeper and larger for positive energies than that
one for negative energies. Thus, the capacity to hold bound states depends
on the sign of the energy and one might expect that the number of
positive-energy levels is greater than the number of negative-energy levels.
By the way, the positive (negative) energy solutions are not to be promptly
identified with the solutions for particles (antiparticles). Rather, whether
it is positive or negative, an energy can be unambiguously identified with a
bounded solution for a particle (antiparticle) only by observing if the
energy level emerges from the upper (lower) continuum. When $g_{1}=0$,
though, the effective cusp potential allows energy levels symmetric about $%
E=0$. The easiness with which the nonminimal vector potential well has bound
states for both signs of $E$ is due to the fact that this kind of potential
couples equally to particles and antiparticles, as has already been
anticipated by the charge-conjugation properties. Condition (\ref{sc4}) can
also be expressed as
\begin{equation}
g_{1}E>-\frac{g_{1}^{2}+g_{2}^{2}}{2}\,V_{0}  \label{sc5}
\end{equation}%
This inequality can be used to achieve the constraint on the potential
parameters as well as the signs of $E$. For $g_{1}>0$ all positive values of
$E$ are allowed but negative values are allowed only if $|E|<\frac{%
g_{1}^{2}+g_{2}^{2}}{2g_{1}}\,V_{0}$. From this last formula, one can see
that only if $V_{0}>2m\frac{g_{1}}{g_{1}^{2}+g_{2}^{2}}$energy levels with $%
E\simeq -m$ will be allowed. When $g_{1}=0$, though, the spectrum can
acquiesce both signs for the energies, regardless the values of $E$, $g_{2}$
and $V_{0}$. In this qualitative context, there is no hint or indication
whether Popov's conjecture is valid. Nevertheless, since the energy levels
for antiparticles must emerge from the lower continuum one can conclude that
the SSWE is a strong-field phenomenon inasmuch as $E\simeq -m$ requires $%
V_{0}>2m$ in the case of a pure minimal coupling attractive for particles in
a nonrelativistic scheme.

Now we move to consider a quantitative treatment of bound-state solutions.
Since the effective potential is even under $x\rightarrow -x$, $\Phi $ can
be expressed as a function of definite parity. Thus, we can concentrate our
attention on the positive half-line and impose boundary conditions on $\Phi $
and $d\Phi /dx$ at $x=0$ and $x=+\infty $. In addition to $\Phi \left(
\infty \right) =0$ in order to ensure normalizability of the DKP spinor, the
boundary conditions can be met in two distinct ways: the even function obeys
the homogeneous Neumann condition at the origin ($d\Phi /dx|_{x=0}=0$)
whereas the odd function obeys the homogeneous Dirichlet condition ($\Phi
\left( 0\right) =0$) .

Defining the dimensionless quantities
\begin{eqnarray}
z &=&z_{0}\exp \left( -\frac{|x|}{\lambda }\right) ,\quad z_{0}=i2\lambda
V_{0}\sqrt{g_{1}^{2}+g_{2}^{2}}  \notag \\
&&  \label{sc6} \\
\kappa &=&\frac{2E\lambda ^{2}g_{1}V_{0}}{z_{0}},\quad \nu =\lambda m\sqrt{1-%
\frac{E^{2}}{m^{2}}}  \notag
\end{eqnarray}%
one obtains%
\begin{equation}
z\frac{d^{2}\Phi }{dz^{2}}+\frac{d\Phi }{dz}+\left( -\frac{z}{4}-\frac{\nu
^{2}}{z}+\kappa \right) \Phi =0  \label{sc7}
\end{equation}

\noindent Let us define $w=z^{1/2}\Phi $ so that $w$ obeys the Whittaker
equation \cite{abr}:
\begin{equation}
\frac{d^{2}w}{dz^{2}}+\left( -\frac{1}{4}+\frac{\kappa }{z}+\frac{1/4-\nu
^{2}}{z^{2}}\right) w=0  \label{sc8}
\end{equation}%
whose solution regular at $z=0$ ($x=\infty $), with

\begin{equation}
a=\frac{1}{2}+\nu -\kappa ,\qquad b=1+2\nu  \label{sc9}
\end{equation}
is written as $w=N\,z^{1/2+\nu }e^{-z/2}M(a,b,z)$. Here $N$ is a
normalization constant and $M(a,b,z)$ is the regular confluent
hypergeometric function \noindent (Kummer's function)%
\begin{equation}
M(a,b,z)=\frac{\Gamma \left( b\right) }{\Gamma \left( a\right) }%
\sum_{n=0}^{\infty }\frac{\Gamma \left( a+n\right) }{\Gamma \left(
b+n\right) }\,\frac{z^{n}}{n!}
\end{equation}%
where $\Gamma \left( z\right) $ is the gamma function. \noindent Thus,
\begin{equation}
\Phi =Nz^{\nu }e^{-z/2}M(a,b,z)  \label{sc10}
\end{equation}

\noindent From Eq. (\ref{sc10}) one can see now that the boundary conditions
at $x=0$ ($z=z_{0}$) imply into%
\begin{equation}
\begin{array}{ll}
\frac{M\left( a+1,b+1,z_{0}\right) }{M\left( a,b,z_{0}\right) }=\frac{%
z_{0}+1-b}{2\frac{a}{b}z_{0}}, & {\text{for even states}} \\
&  \\
M(a,b,z_{0})=0, & {\text{for odd states}}%
\end{array}
\label{sc11}
\end{equation}%
Since the DKP energies are dependent on $a$ and $b$ via $\kappa $ and $\nu $%
, it follows that Eq. (\ref{sc11}) is a quantization condition. The allowed
values for the parameters $a$ and $b$, and $E$ as an immediate consequence,
are determined by solving Eq. (\ref{sc11}). Due to the presence of $\kappa $
in the definition of $a$ one can expect an asymmetry in the spectrum so that
the presence of each sign of energy depends, of course, on the relative
strength between $g_{1}$ and $g_{2}$. For the case of a pure nonminimal
vector potential ($g_{1}=0$), though, one has $\kappa =0$ \ and the
negative- and positive-energy levels are disposed symmetrically about $E=0$,
as commented before, and there are as many positive-energy levels as
negative ones. This particular case allows a simpler mathematical treatment
as can be seen in Appendix A. Although the quantization condition has no
closed form expressions in terms of simpler functions, the exact computation
of the allowed eigenenergies can be done easily with a root-finding
procedure of a symbolic algebra program. Proceeding in this way, the whole
bound-state spectrum is found. The DKP energies are plotted in Figure 1, 2
and 3 for the lowest bound-state solutions as a function of $V_{0}$ for two
different values of $\lambda $ ($\lambda _{c}=m^{-1}$ is the Compton
wavelength). The energies for $g_{1}<0$ and $g_{2}<0$ can be obtained by
using the charge-symmetries mentioned before, viz. the spectrum is invariant
\ under $g_{2}\rightarrow -g_{2}$, and $E\rightarrow -E$ when $%
g_{1}\rightarrow -g_{1}$. For large $\lambda $ and small $V_{0}$, the
spectrum consists of a finite set of energy levels of alternate parities and
the energy level corresponding to the ground-state solution ($\Phi $ even)
of particles (for $g_{1}>0$) always makes its appearance, as it happens in a
nonrelativistic framework. Surprisingly, as $V_{0}$ increases the number of
bound-state solutions can decrease and one might find an odd-parity
ground-state solution, or no solution at all. Indeed, the situation is more
complicated and the existing correlation between the number of bound-state
solutions and the depth of the effective potential well in the
nonrelativistic theory does not verify for a strong potential. In the case
of a pure minimal coupling, as illustrated in Figure 1, particle levels
appear in the spectrum and those levels tend to sink at the continuum of
negative energies for large $\lambda $. As $\lambda $ decreases, though, a
new branch of solutions corresponding to antiparticle levels begin to emerge
from the continuum of negative energies and coalesce with the particle
levels. This is the signature of the SSWE. In Figure 2 one can see the
tendency of the spectrum to be symmetrical about $E=0$. In Figure 3, for the
case of a pure nonminimal coupling, that symmetry shows itself perfect.

\section{Conclusions}

In summary, we have succeed in the proposal of searching the solution for a
cusp potential with the DKP equation. An opportunity was given to analyze
some aspects of the DKP equation which would not be feasibly only with the
cases already approached in the literature. Thus, the use of the mixing of
minimal and nonminimal vector Lorentz structures for other kinds of
potentials, primarily because nonminimal vector couplings have no
counterpart in the Klein-Gordon and Proca theories, may lead to a better
understanding of the DKP equation and its solutions. In the case of a pure
minimal coupling ($g_{2}=0$) the DKP equation reduces to the Klein-Gordon
and Proca equations and our results are in accordance with those found in
the literature for the Klein-Gordon case (see e.g. \cite{vil2}). In
particular, the Popov conjecture about the SSWE is supported as a
short-range phenomenon for spin-0 particles as well as for spin-1 particles.

Finally, one very important point to note is that the matrix potential $%
i[P,\beta ^{\mu }]A_{\mu }^{\left( 2\right) }$ in (\ref{dkp2}) leads to a
conserved four-current but the same does not happen if instead of the matrix
$i[P,\beta ^{\mu }]$ one uses either $P\beta ^{\mu }$ or $\beta ^{\mu }P$,
as in \cite{cla1}-\cite{cla2}, even though the linear forms constructed from
those matrices behave as true Lorentz vectors.

\bigskip

\bigskip

\bigskip

\bigskip

\bigskip

\noindent {\textbf{Acknowledgments}}

The author is indebted to anonymous referees for their criticisms. This work
was supported in part by means of funds provided by Conselho Nacional de
Desenvolvimento Cient\'{\i}fico e Tecnol\'{o}gico (CNPq).

\newpage

\noindent {\large {\textbf{Appendix A}}}

In the particular case of a pure nonminimal vector potential one can take
advantage of the relation expressing the Bessel function of the first kind
and order $\nu $ in terms of the hypergeometric function \cite{abr}
\begin{equation}
J_{\nu }\left( z\right) =\frac{\left( \frac{1}{2}z\right) ^{\nu }e^{-iz}}{%
\Gamma \left( \nu +1\right) }\,M\left( \nu +\frac{1}{2},2\nu +1,2iz\right)
\label{sc12}
\end{equation}%
to write
\begin{equation}
\Phi =N_{\nu }\,J_{\nu }(y)  \label{sc13}
\end{equation}%
\noindent where $N_{\nu }$ is a normalization constant and $y=-iz/2$. The
boundary conditions at $x=0$ ($y=y_{0}$) imply that
\begin{equation}
\begin{array}{ll}
\frac{dJ_{\nu }(y)}{dy}|_{y=y_{0}}=0, & {\text{for even states}} \\
&  \\
J_{\nu }(y_{0})=0, & \text{for odd states}%
\end{array}
\label{sc14}
\end{equation}

\noindent The oscillatory character of the Bessel function and the finite
range for $y$ ($0<y\leq y_{0}$) imply that there is a finite number of
discrete DKP energies. The roots of $J_{\nu }(y)$ and $J_{\nu }^{\prime }(y)$
are listed in tables of Bessel functions only for a few special values of $%
\nu $. A bit of time and effort can be saved in the numerical calculation of
the roots of $J_{\nu }^{\prime }(y)$ if one uses the recurrence relation $%
J_{\nu -1}-J_{\nu +1}=2J_{\nu }^{\prime }$, in such a manner that the
quantization condition for even states translates into $J_{\nu
+1}(y_{0})=J_{\nu -1}(y_{0})$.

\begin{figure}[th]
\begin{center}
\subfigure[$\lambda =\lambda_{c}$]{
\includegraphics[width=6.2cm]{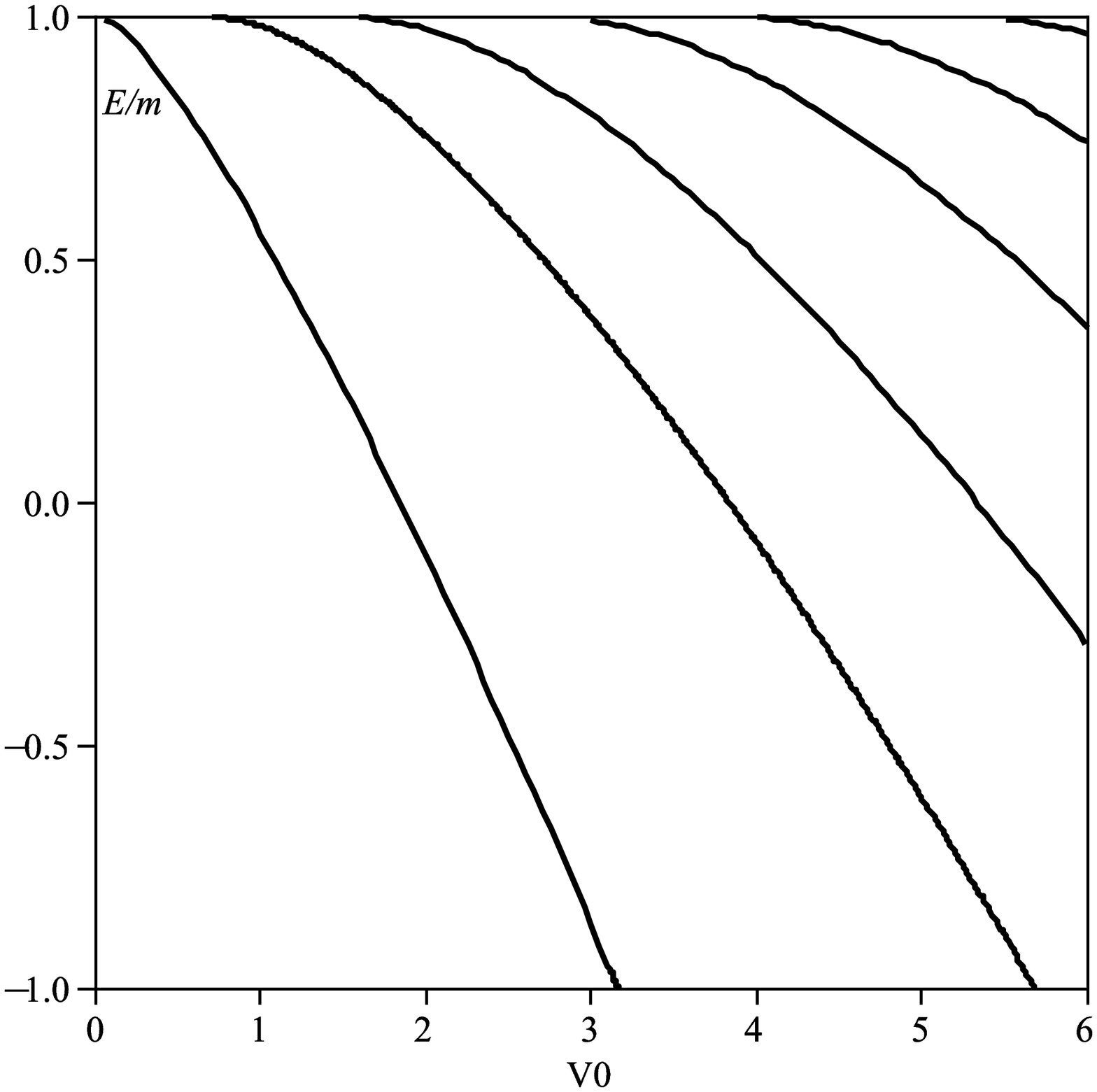}
\label{fig:Fig1a}
}
\subfigure[$\lambda =\lambda_{c}/10$]{
\includegraphics[width=6.2cm]{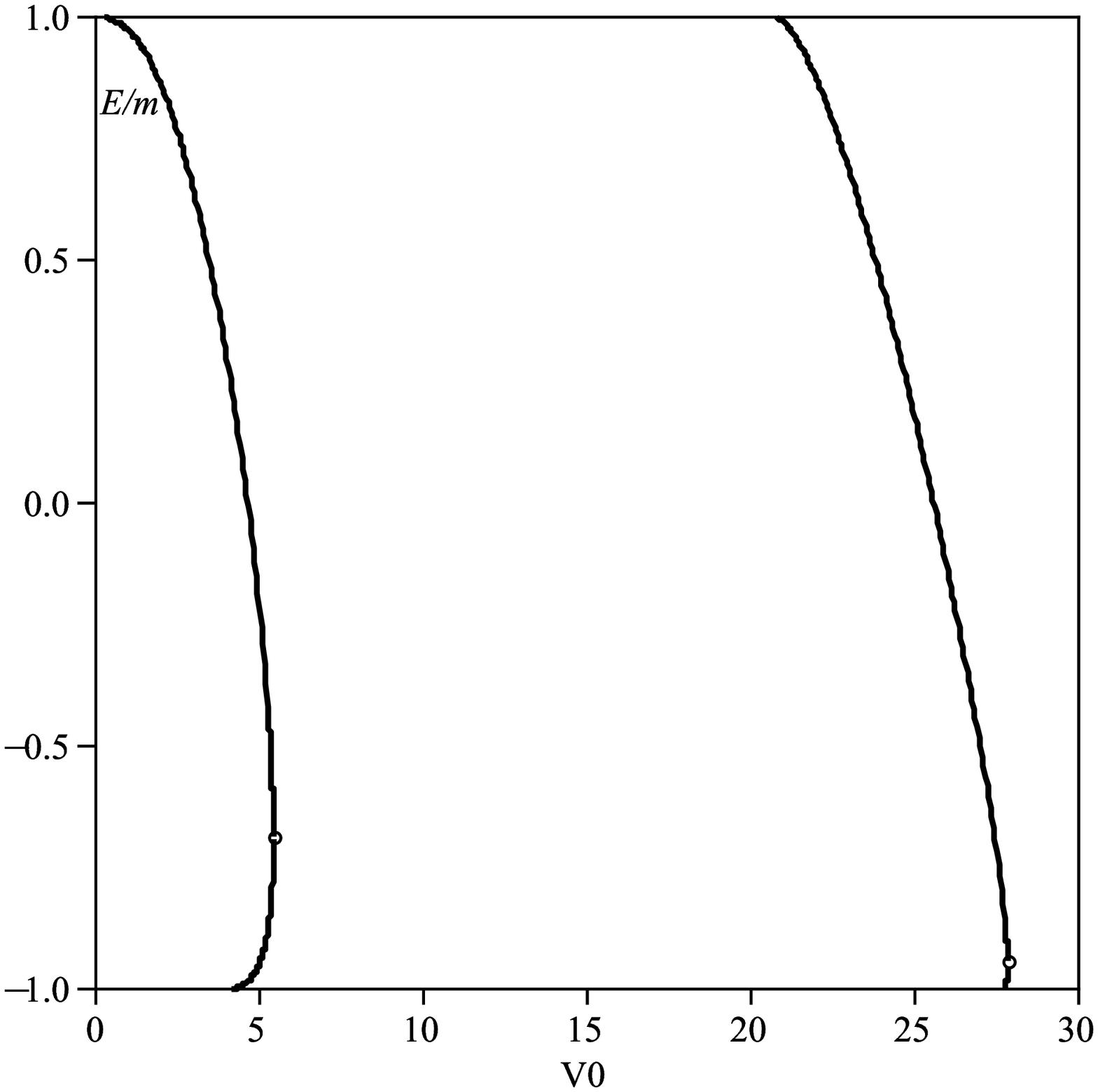}
\label{fig:Fig1b}
}
\end{center}
\caption{Energies for the first bound states as a function of $V_{0}$ for $%
g_{1}=1$ and $g_{2}=0$.}
\end{figure}

\begin{figure}[th]
\begin{center}
\subfigure[$\lambda =\lambda_{c}$]{
\includegraphics[width=6.2cm]{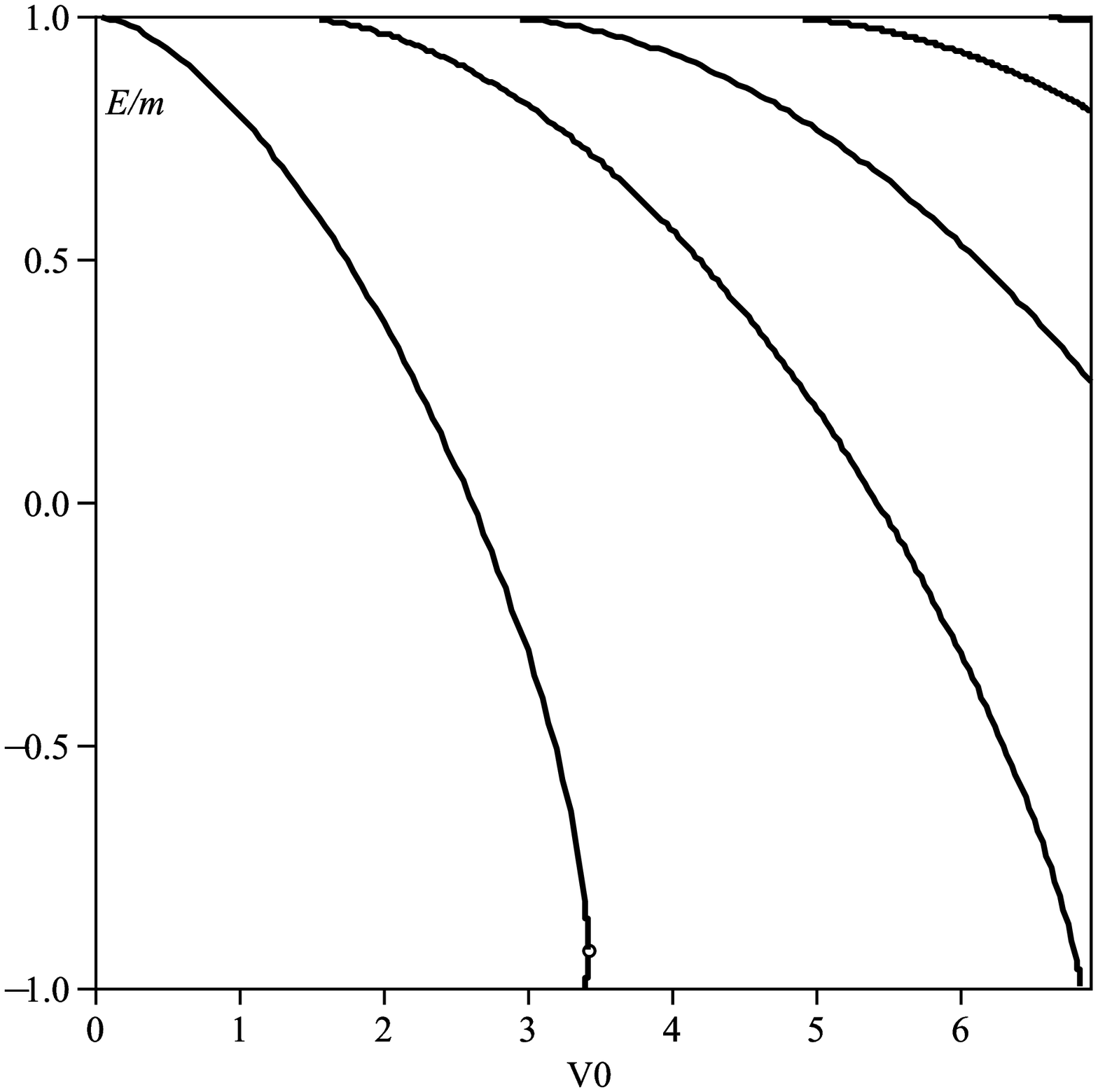}
\label{fig:Fig2a}
}
\subfigure[$\lambda =\lambda_{c}/10$]{
\includegraphics[width=6.2cm]{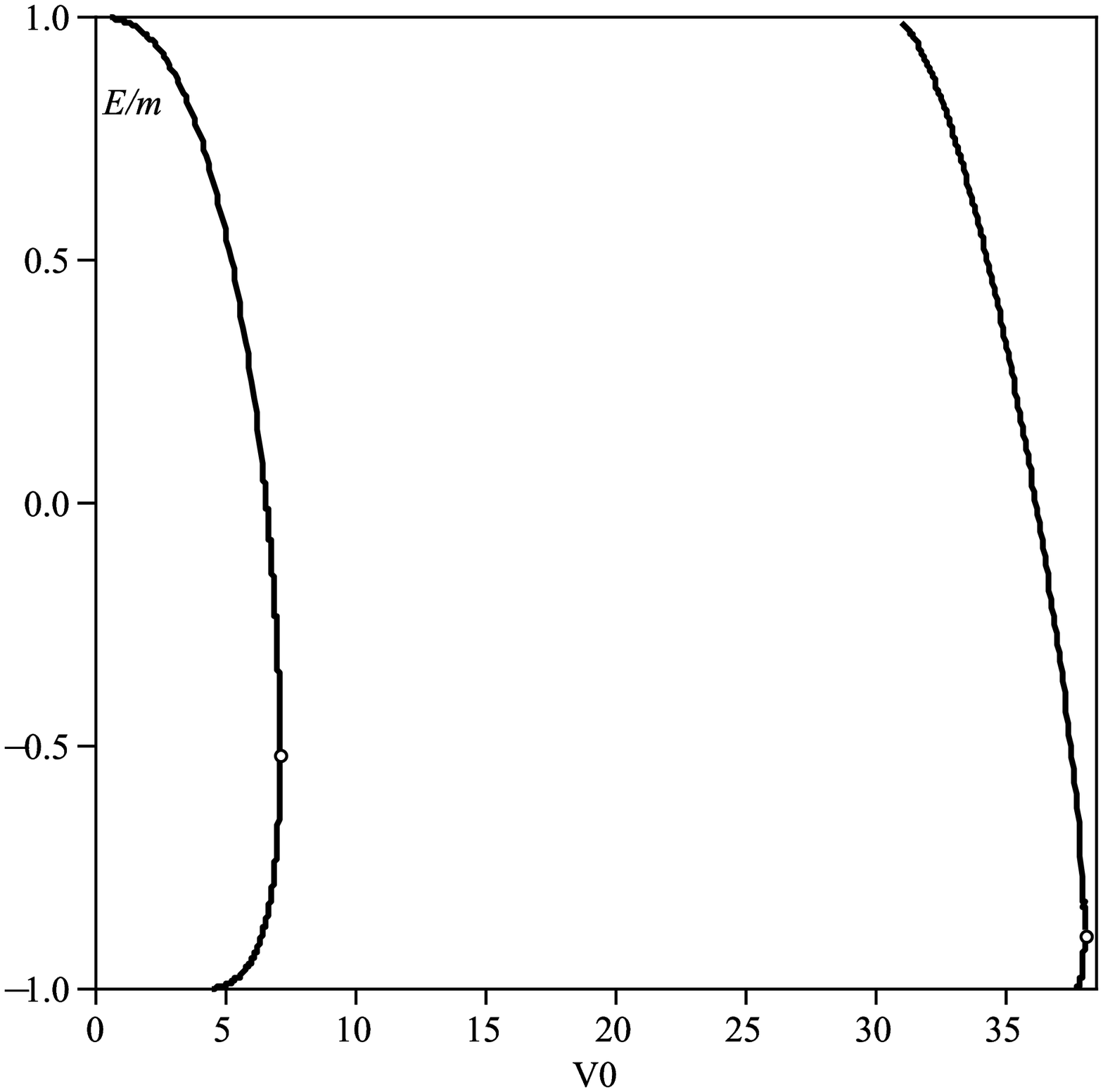}
\label{fig:Fig2b}
}
\end{center}
\caption{Energies for the first bound states as a function of $V_{0}$ for $%
g_{1}=g_{2}=1/2$.}
\end{figure}

\begin{figure}[th]
\begin{center}
\subfigure[$\lambda =\lambda_{c}$]{
\includegraphics[width=6.2cm]{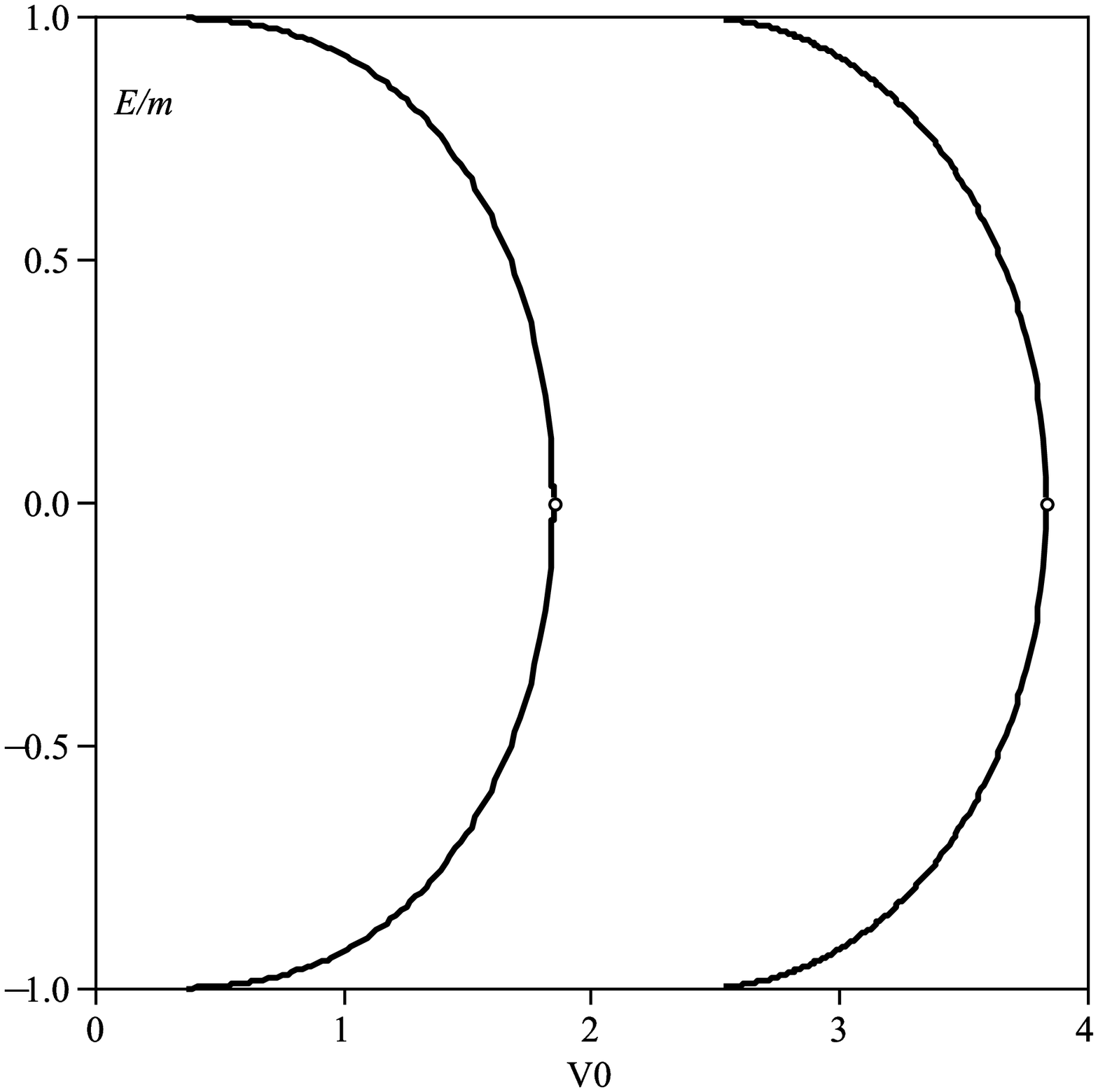}
\label{fig:Fig3a}
}
\subfigure[$\lambda =\lambda_{c}/10$]{
\includegraphics[width=6.2cm]{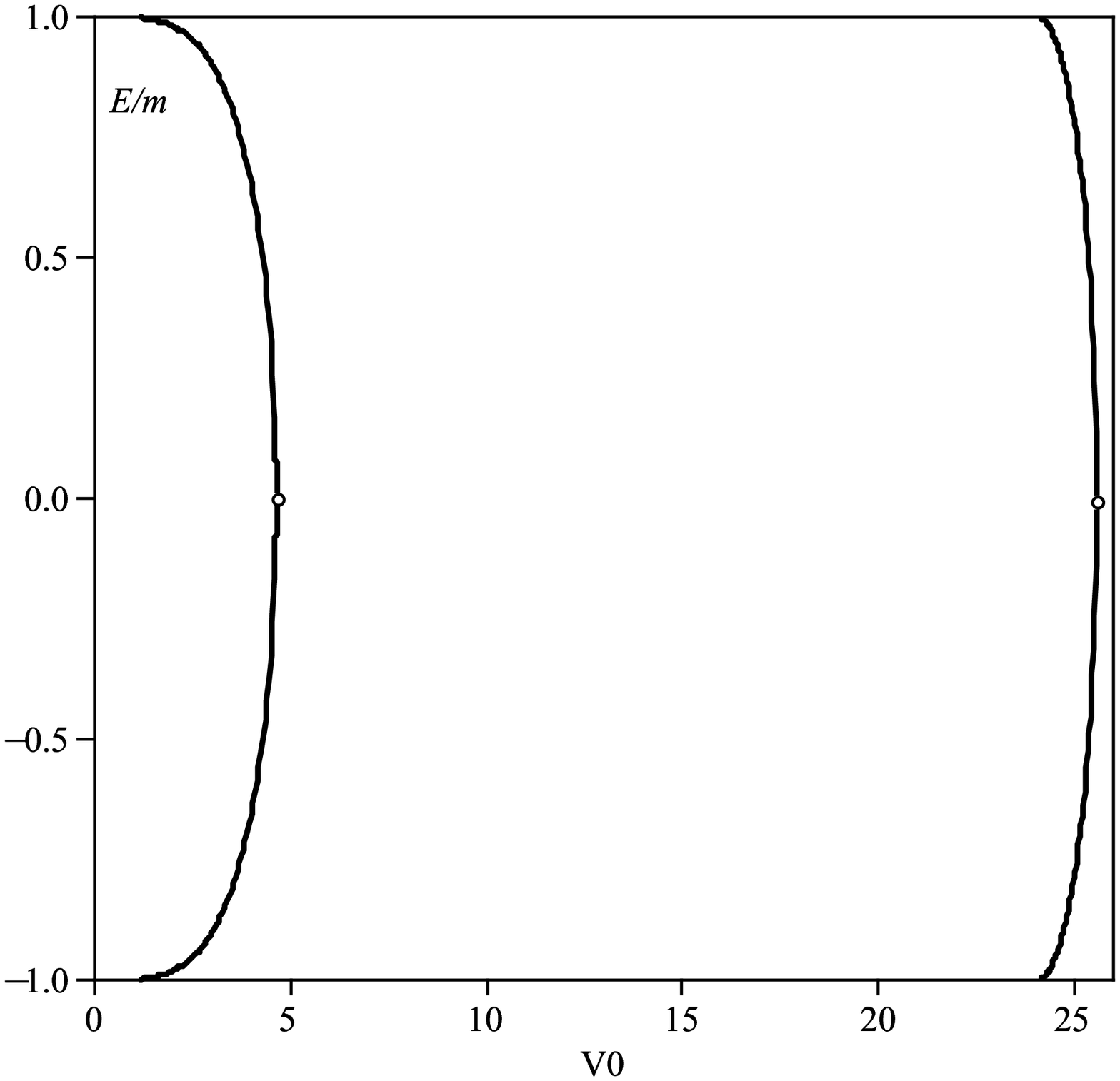}
\label{fig:Fig3b}
}
\end{center}
\caption{Energies for the first bound states as a function of $V_{0}$ for $%
g_{1}=0$ and $g_{2}=1$.}
\end{figure}

\end{document}